# Lines of Bound States in the Continuum in a Phononic Crystal Slab

Lin Yang[1*], Riyi Zheng[2*], Sheng Zhang[3*], Wenshuai Zhang[1], Qiujiao Du[1], Pai Peng[1], Ziyu Wang[3†], Manzhu Ke[3‡], Xueqin Huang[2§], and Fengming Liu[1∥]

[1]School of Mathematics and Physics, China University of Geosciences, Wuhan 430074, China

[2]School of Physics and Optoelectronics, South China University of Technology, Guangzhou, Guangdong 510640, China

[3]Key Laboratory of Artificial Micro- and Nanostructures of Ministry of Education and School of Physics and Technology, Wuhan University, Wuhan 430072, China

[*]These authors contributed equally to this work.

[†]zywang@whu.edu.cn

[‡]mzke@whu.edu.cn

[§]phxqhuang@scut.edu.cn

[∥]liufm@cug.edu.cn

**ABSTRACT**

We demonstrate that bound states in the continuum (BICs) form continuous lines along high-symmetry directions of momentum space in a simple phononic crystal slab. Contrary to common sense, these BICs are symmetry-protected (SP) BICs not only at the center of the Brillouin zone ($\Gamma$ point) but also off the $\Gamma$ point. We utilize numerical simulations, a group theory method, and a mode expansion method to comprehensively understand the formation of the BICs. It is revealed that these BICs correspond to phase singularity lines of far-field radiation, and the phase winding number can be defined as a topological index. This makes the BICs topologically protected and robust to any parameter variation that maintains periodicity and rotational symmetry. Finally, the generation of the BICs lines is experimentally demonstrated.

Bound states in the continuum are eigenmodes whose energies lie in the continuous spectrum of radiation states. Since BICs was first mathematically proposed by Von Neumann and Wigner in the framework of quantum mechanics [1], the concept has been extended to classical wave systems [2-32]. Their unique characteristics in terms of strong localization and high-quality factor lead to extensive applications in lasers, sensors, filters, and wavefront control [33]. Various mechanisms [2], such as symmetry protection [3-10,27-30], parameter tuning [11-22,31,32], and environment engineering [23-26], have been proposed to achieve BICs. Among them, symmetry protection is the most straightforward mechanism. It is predictable from a group theory perspective and, as long as symmetry is preserved, no tuning of system parameters is required to achieve it. Thus, much of the current research around BICs focuses on SP BICs in photonic crystal slabs [4-9].

For acoustic systems, a SP BICs in a "plate-in-waveguide" system was first observed in the 1960s [27,28]. Later, different types of acoustic BICs, including Friedrich–Wintgen BICs, Fabry–Perot BICs and mirror-induced BICs have been proposed [29-32,34-41]. However, most previous acoustic BICs systems consisted of resonators attached to one-dimensional (1D) waveguides. Few studies have focused on BICs in phononic crystal slabs [42,43]. It has been demonstrated that photonic BICs in photonic crystal slabs are vortex polarization singularities in momentum space, which exhibits their intriguing topological features [14-22]. Since phononic crystal slabs are the acoustic counterpart of photonic crystal slabs, the reasonable questions to consider are, what types of BICs can phononic crystal slabs support? Can acoustic BICs have a topological description similar to photonic BICs in momentum space?

It was well studied that spoof surface acoustic waves (SSAWs) can be supported by a rigid slab drilled with periodic subwavelength hole array [44-48]. However, SSAWs are conventional bound states that cannot couple to the radiation channel because they only exist below the sound line. In this work, we demonstrate that, in addition to SSAWs, BICs can also appear in the similar structures with widened holes. Contrary to the common sense that SP BICs appear at the Γ point and accidental BICs appear off the Γ point [4-22], the BICs in our structures are SP BICs both at and off the

Γ point. They form continuous lines along high-symmetry directions of momentum space. The formation of our SP BICs can be understood comprehensively by utilize group theory [49-51] and mode expansion methods [52-55]. It is found that the destructive interaction between the radiations from the $+1$st and $-1$st-order waveguide modes in the hole leads to the realization of the BICs. We further reveal that the BICs lines correspond to phase singularity lines of radiation field. In 3D momentum-frequency space, a conserved topological charge can be defined by the phase winding numbers, which makes the BICs topologically protected and robust to any parameter variation maintaining periodicity and rotational symmetry. Note that in contrast to the idealized pure magnetic octupole lattice used to realize BICs lines [56,57], our structure is simple, realistic and does not require a radiative environment design [23-26].

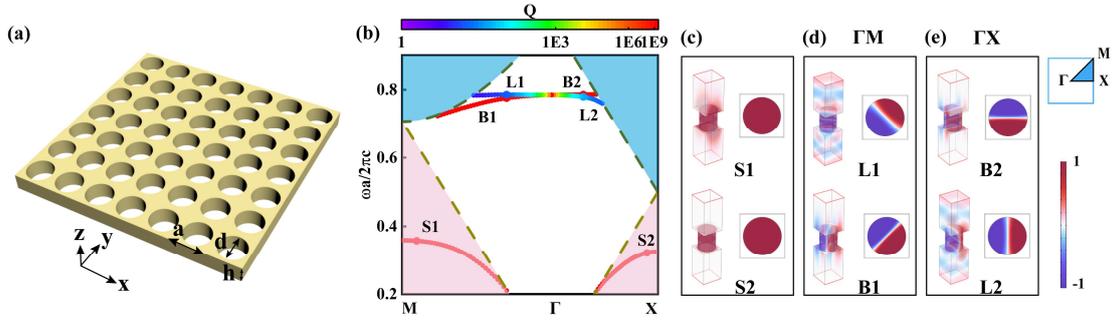

FIG. 1. Band structures, quality factor and pressure fields of the eigenmodes for the phononic slab. (a) Schematic view of a phononic slab. (b) The numerically calculated band structures and quality factor of the SSAWs, leaky modes, and BICs. (c) Pressure fields of the eigenmodes for the SSAWs bands. (d), (e) Pressure fields of the eigenmodes for the leaky and BICs bands along the ΓM and ΓX directions, respectively.

We start by considering a rigid slab drilled with square lattice array of cylindrical holes (Fig. 1(a)). The structure is characterized by lattice constant $a$, cylinder diameter $d = 0.85a$, and slab thickness $h = 1.0a$. We first investigate the phononic slab using the finite-element method (see Supplemental Material [55]). Fig. 1(b) shows the numerically calculated band structures of the SSAWs, leaky modes, and BICs for the structure. In addition to the SSAWs below the sound line, two dispersion curves appear above the sound line and below the first Bragg diffraction limit. Remarkably, at the Γ point, the two states are degenerate and both have infinite high $Q$ factors; while off the

Γ point, only one of the two dispersion curves has infinite high $Q$ factors. It is quite novel that the BICs form a line because BICs generally appear at discrete $k$ points as reported in previous works [2-22]. To understand the formation mechanism of the BICs, the eigen pressure fields of the SSAWs, leaky, and BICs bands are calculated and shown in Fig. 1(c-e). Unlike SSAWs which involves only the 0th-order hole mode, the leaky and the BICs bands also include the ±1st-order hole modes. From the calculated pressure fields, we can expect that the symmetry of the eigenmodes plays a key role in the formation of the BICs. According to the group theory analysis of photonic crystal slabs [50,51], any eigenmode is an irreducible representation (irrep) of the corresponding point group. Our structure has $D_{4h}$ symmetry, which is a direct product of the $C_{4v}$ and $C_{1h}$ point groups: $D_{4h} = C_{4v} \times C_{1h}$ [50,51]. $C_{1h}$ consists of the identity operation and the mirror reflection by the *x-y* plane, $\sigma_z$. Any eigenmode of the structure should be symmetric ($\sigma_z = 1$) or antisymmetric ($\sigma_z = -1$) about the *x-y* plane. Thus, the radiative decay rate of a resonance toward the top of the structures is always the same as that toward the bottom, and both rates are reduced to zero at BICs. In order to avoid complexity, we have used $C_{4v}$ instead of $D_{4h}$ symmetry as the structure has $\sigma_z$ symmetry.

Here, the point group at the Γ point is the $C_{4v}$ group, which has four 1D irreps $A_1$, $A_2$, $B_1$, $B_2$ and one 2D irrep $E$. For points on the segments ΓX and ΓM, the corresponding point group is the $C_{1h}$ group, which has two 1D irreps $A$ and $B$. The two eigenmodes are degenerate at the Γ point, which correspond to the 2D irreps $E$. The $E$ modes are odd with respect to rotation about the *z* axis by $C_2$. It is well known that a plane wave in cylindrical coordinates can be expanded as [54,55],

$$p = p_0[J_0(Q_0 r) + 2\sum_{m=1}^{\infty} i^m \cos(m\varphi) J_m(Q_0 r)]\exp(iq_0 z). \tag{1}$$

As shown in Eq. (1), an external propagating plane wave does not have circumferential variation when radiating vertically (only $J_0(Q_0 r)$ items left), which means that the radiation channel is even with respect to $C_2$. Thus, the $E$ modes cannot radiate because of this symmetry mismatch, i.e., they are SP BICs. Away from the Γ point, there are two separated bands due to the interaction of the hole modes. Considering the compatibility relations, the *E* modes at the $\Gamma$ point connect to the *A* and *B* modes along high-symmetry directions ($\Gamma X$ and $\Gamma M$). The *A* modes are symmetric under the mirror reflection, whereas the *B* modes are antisymmetric under the same mirror reflection. The radiation channel (Eq. (1)) is symmetric under the mirror reflection because it is a cosine function of azimuth angle $\varphi$. Therefore, the *A* modes are leaky modes while the *B* modes are SP BICs. The numerically calculated pressure fields confirm the group theory analysis.

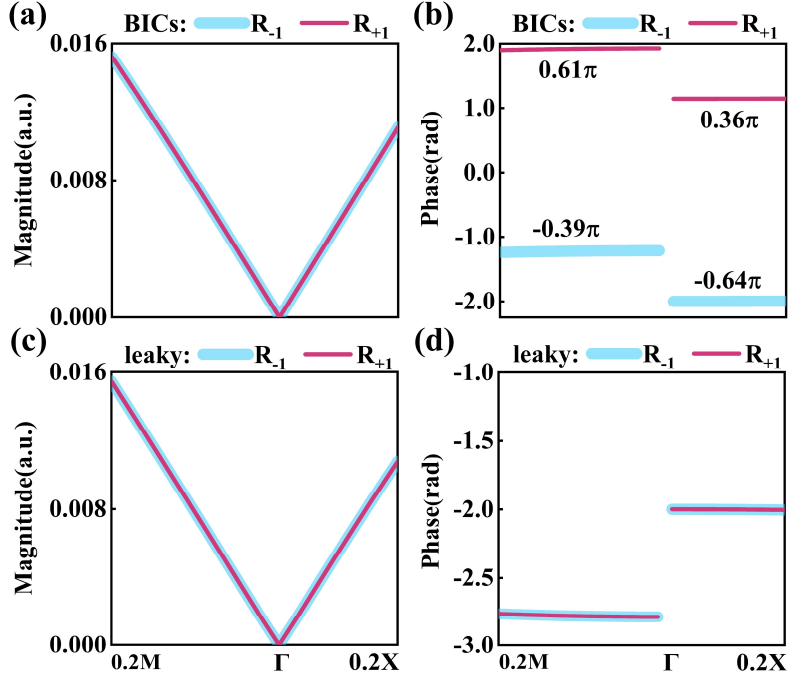

FIG. 2. An analytical method for understanding the formation of the SP BICs. (a), (b) The magnitudes and phases of the radiations from the $\pm$ 1st-order hole modes for the BICs band. (c), (d) The magnitudes and phases of the radiations from the $\pm$ 1st-order hole modes for the leaky band.

Next, to gain a deeper understanding of the generation of the BICs, a mode expansion method is exploited to calculate the complex frequency, pressure fields and coupling coefficients of the eigenmodes. (A detailed derivation of the mode expansion method can be found in Supplemental Material [55]). The pressure fields above and below the slab are expanded as the sum of diffraction plane waves, respectively

$$p_1 = 2\exp(i\mathbf{Q_0}\cdot\mathbf{r}_\parallel)\cos(q_0 z) + \sum_{\mathbf{G}}\beta_{\mathbf{G}}^+ \exp(i(\mathbf{Q_G}\cdot\mathbf{r}_\parallel + q_{\mathbf{G}}z)), \qquad (2)$$

$$p_3 = \sum_{\mathbf{G}} \beta_{\mathbf{G}}^- \exp(i(\mathbf{Q}_{\mathbf{G}} \cdot \mathbf{r}_\parallel - q_{\mathbf{G}}(z+h))). \tag{3}$$

Inside the holes, the pressure field can be expanded as the linear combination of the waveguide modes of different orders,

$$p_2 = \sum_{m=-\infty}^{\infty} J_m(Q_m r) \exp(im\varphi) \psi_m^+(z), \tag{4}$$

with

$$\psi_m^\pm(z) = [\alpha_m^+ \exp(iq_m(z+h)) \pm \alpha_m^- \exp(-iq_m z)]. \tag{5}$$

Here, $\mathbf{k}_0 = (\mathbf{Q}_0, q_0)$ is the incident wave vector in free space, $\mathbf{r}_\parallel = (x, y) = (r, \varphi)$; $\mathbf{Q}_{\mathbf{G}} = \mathbf{Q}_0 + \mathbf{G}$, $\mathbf{G}$ is the reciprocal lattice vector of square lattice, $q_{\mathbf{G}} = \sqrt{\mathbf{k}_0^2 - \mathbf{Q}_{\mathbf{G}}^2}$ and $q_m = \sqrt{\mathbf{k}_0^2 - Q_m^2}$ are the wavenumbers in the $z$ direction satisfying the Helmholtz equation outside and inside the hole; $J_m$ denotes the $m$th order Bessel function, $dJ_m(Q_m r)/dr \big|_{r=r_0} = 0$, $r_0$ is the radius of the hole, and $Q_m = x_m/r$ is the radial wave vector, with $x_m$ being the $m$th zero of $J_m$. $\beta_{\mathbf{G}}^\pm$ corresponds to the coupling coefficients between the hole modes and the diffraction plane waves. By requiring the continuity of particle velocity and sound pressure at the hole openings, vanishing particle velocity elsewhere on the interfaces, a set of linear equations for the expansion coefficients $\alpha_m^\pm$ can be obtained,

$$\begin{bmatrix} \mathbf{D}_{11} & \mathbf{D}_{12} \\ \mathbf{D}_{21} & \mathbf{D}_{22} \end{bmatrix} \begin{bmatrix} \alpha_m^+ \\ \alpha_m^- \end{bmatrix} = \begin{bmatrix} 2(I_{m'}(\mathbf{Q}_0))^* \\ 0 \end{bmatrix}, \tag{6}$$

where $2(I_{m'}(\mathbf{Q}_0))^*$ depends on the overlap integral between the incident wave and the hole modes. The dispersion curves of the structure can be obtained by setting $2(I_{m'}(\mathbf{Q}_0))^* = 0$, thus solving the secular equation $\det|\mathbf{D}_{ij}| = 0$, and the corresponding coefficients $\alpha_m^\pm$ can be used to calculate the eigen fields.

The band structure and eigen pressure fields of the phononic slab calculated using the mode expansion method are presented in Fig. S3 (see Supplemental Material [55]). The good agreement between the analytical and numerical results validates the effectiveness of the mode expansion method. Note that the coupling coefficient $\beta_G^\pm$ can be regarded as the complex amplitude of the radiation waves as shown in Eqs. (2) and (3), so we expect that the occurrence of the BICs can be explained by examining these coupling coefficients. The coupling coefficients $\beta_G^\pm$ can be determined as,

$$\beta_G^+ = \frac{1}{S}\sum_m \frac{q_m}{q_G}\psi_m^-(0)I_m(\mathbf{Q_G}), \tag{7}$$

$$\beta_G^- = -\frac{1}{S}\sum_m \frac{q_m}{q_G}\psi_m^-(-h)I_m(\mathbf{Q_G}), \tag{8}$$

where $S$ is the unit cell area and

$$I_m(\mathbf{Q_G}) = \int_0^{2\pi}\int_0^{r_0} J_m(Q_m r)e^{i(m\varphi - \mathbf{Q_G}\cdot\mathbf{r}_\parallel)}r\,dr\,d\varphi \tag{9}$$

represents the overlap integral between outgoing plane waves and the hole modes. For the frequencies below the first diffraction limit, only the $\mathbf{G} = 0$ term gives a nonzero contribution to the radiation waves. and due to the up–down mirror symmetry of the structure, the radiation waves on the top and bottom of the slab are equal, therefore only the coefficient $\beta_0^+$ has been calculated. According to Eq. (7), each hole mode contributes to the radiation. In the calculation, the 0th-order and the $\pm$ 1st-order hole modes are considered since these modes are propagating modes within the frequency range of interest. We find that the radiation from the 0th-order hole mode is two orders of magnitude smaller than the radiations from the $\pm$ 1st-order hole modes, thus the contribution of the 0th-order hole mode can be ignored. For the sake of brevity hereafter $\beta_0^+$ is denoted as $R$, that is, $R \equiv \beta_0^+$. When only the $\pm$ 1st-order hole modes are considered, Eq. (7) can be conveniently rewritten as $R = R_{+1} + R_{-1}$. As shown in Fig. 2(a), for the BICs band, the magnitudes of the radiation from the two hole modes are equal everywhere along the high-symmetry directions. Meanwhile, the phase differences between $R_{+1}$ and $R_{-1}$ are shown in Fig. 2(b). It can be seen that $R_{+1}$ is $\pi$ radians out of phase with $R_{-1}$. The two hole modes interfere destructively with each other so that the total radiation $R$ drops to zero. For the leaky band, the magnitudes of the radiation from the two hole modes are also equal everywhere along the high-symmetry directions, but this time the radiations from the two hole modes are in phase as shown in Fig. 2(c) and 2(d). A nonzero total radiation $R$ indicates that the eigenmodes of leaky band can couple with the radiation continuum. We also note that the radiation at the $\Gamma$ point is a special case for both bands. Both $R_{+1}$ and $R_{-1}$ are zero, thus the phase of them cannot be determined. This can be explained by Eq. (9). At normal incidence, the integral $I_{\pm 1}$ for the $\pm$ 1st-order hole modes are zero, which indicates that the hole modes cannot couple with the radiation continuum. In addition, we present in Supplemental Material [55] the complex amplitude of the radiation $R_{\pm 1}$ for any general direction in momentum space. The unequal radiation magnitudes prevent the two hole modes from interfering destructively with each other, thus the BICs only exist along the high-symmetry directions.

Recently, photonic BICs in photonic crystal slabs have been proved to be polarization singularities of far-field radiation in 2D momentum space [14-22]. They carry conserved and quantized topological charges, defined by the winding numbers of the polarization vectors, which ensure their robust existence. Obviously, the viewpoint of BICs corresponding to polarization singularities cannot be extended to acoustic systems because sound waves are scalar waves without polarization vectors. However, note that the radiation sound field $R$ is a complex number and can be expressed as

$$R = V \exp(i\Omega), \tag{10}$$

where $V$ and $\Omega$ are the amplitude and phase of the radiation sound field, respectively. The phase $\Omega$ is ill defined if $V = 0$, which indicates the vanishing of radiation amplitude is equivalent to a singular value of its phase. In other words, acoustic BICs intrinsically correspond to phase singularities of radiation sound field.

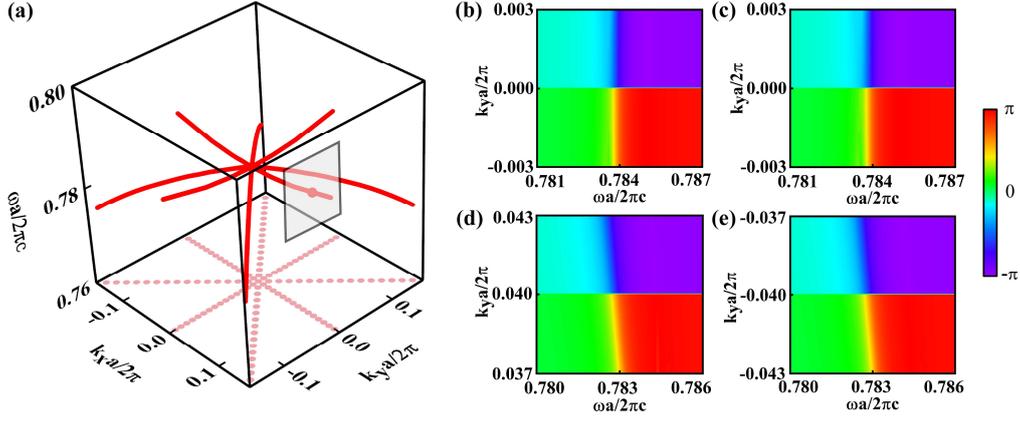

FIG. 3. Topological description of the BICs lines. (a) Dislocation lines in 3D momentum-frequency space. (b)-(e) The phase distributions of the radiation fields in the planes intersected with the dislocation lines at ($\omega = 0.784$, $k_x = 0.05$, $k_y = 0$), ($\omega = 0.784$, $k_x = -0.05$, $k_y = 0$), ($\omega = 0.783$, $k_x = 0.04$, $k_y = 0.04$), and ($\omega = 0.783$, $k_x = -0.04$, $k_y = -0.04$), respectively.

As shown in Fig. 3(a), these singularities compose dislocation lines in 3D momentum-frequency space. The common method of defining topological charges of discrete BICs in 2D momentum space is not applicable here because the BICs form continuous lines in the 2D momentum plane (dotted lines in Fig. 3(a)), whereas a closed path within the same plane will intersect the BICs lines which results in zero phase integral on the closed path, i.e. zero topological charge [57]. Therefore, it is necessary to define the topological charge of the dislocation lines in the full 3D momentum-frequency space. It is known that the phase of a wavefield changes an integer multiple of $2\pi$ when encircling around the dislocation line [58-60]. To quantitatively specify the dislocation strength of phase winding, we consider a counterclockwise closed loop $C$ around the dislocation line. An integral of phase on $C$ defines a topological charge as

$$S = \frac{1}{2\pi}\oint_C d\Omega = \frac{1}{2\pi}\oint_C \nabla\Omega d\mathbf{k}, \tag{11}$$

here $\mathbf{k}$ represents the variable in the 3D momentum-frequency space ($\omega, k_x, k_y$). The dislocation strength is conserved as $C$ slides up and down the dislocation lines [58-60]. Fig. 3(b-e) shows the phase distribution of the radiation field (the calculation method of these radiation fields can be found in Supplemental Material [55]) in the cross-sectional plane of the loop $C$ when the loop $C$ moves along the dislocation lines. The cross-sectional plane is pierced by the dislocation line and the intersection point represents the BICs. We have arbitrarily selected four BICs points, each of which corresponds to a phase vortex in the phase field. By using Eq. (11), the topological charge can be calculated as $S=+1$ for all four BICs points, which confirms that the dislocation strength is a conserved quantity. Due to the existence of conserved topological charge, the BICs lines are topologically protected. The band structures for the phononic crystal slabs with different geometry parameters are shown in Supplemental Material [55], which confirm that the robust existence of the BICs lines.

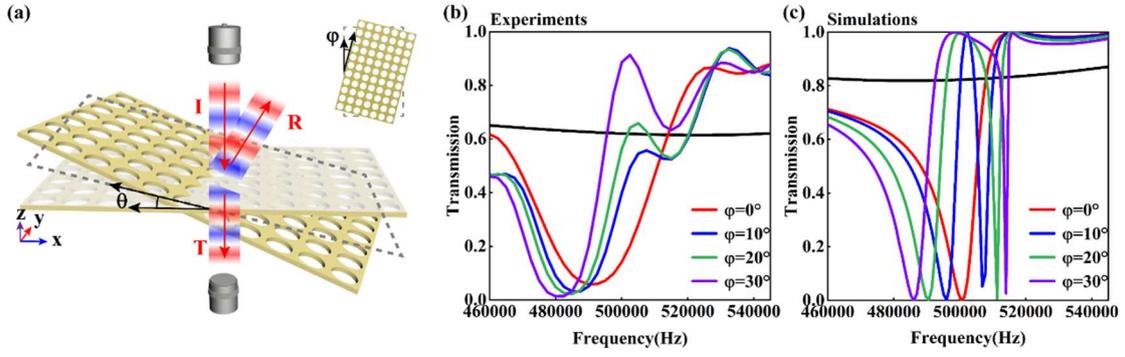

FIG. 4. Experimental demonstration of the BICs. (a) The schematic of the experimental setup. (b) The experimentally measured transmission spectra versus the frequency for the same angle $\theta=16°$ while different angles $\varphi=0°,10°,20°,30°$, respectively. (c) The simulated transmission spectra for the same incidence angles. The black lines in the two figures indicate the results of normal incidence ($\theta=0°,\varphi=0°$).

Finally, we experimentally demonstrate the generation of BICs lines by measuring the transmission of the phononic crystal slab. The schematic of the experimental setup is shown in Fig. 4(a) and the details of the experimental setup are given in Supplemental Material [55]. Fig. 4(b) shows the measured transmission spectra versus the frequency for the same angle $\theta=16°$ while different angles $\varphi=0°,10°,20°,30°$, respectively, which are in good agreement with the simulated results (Fig. 4(c)). The black lines in the two figures indicate the results of normal incidence ($\theta=0°,\varphi=0°$). Since the two states of the structure are BICs at the $\Gamma$ point, they are hidden and do not give any signature in the transmission spectra. For $\theta$ deviates from $0°$, as shown by the red lines only the asymmetric Fano resonance related to the leaky modes appears while the off-$\Gamma$ BICs ($\varphi = 0°$) remains hidden. It is worth noting that the BICs will transform into quasi-BICs along general directions of momentum space [55]. Thus, the BICs can be revealed in the transmission measurement by sweeping the angle $\varphi$ [26]. As shown in Fig. 4(b) and 4(c), another asymmetric Fano resonance associated with the quasi-BICs appears in the transmission spectra when $\varphi$ deviates from $0°$.

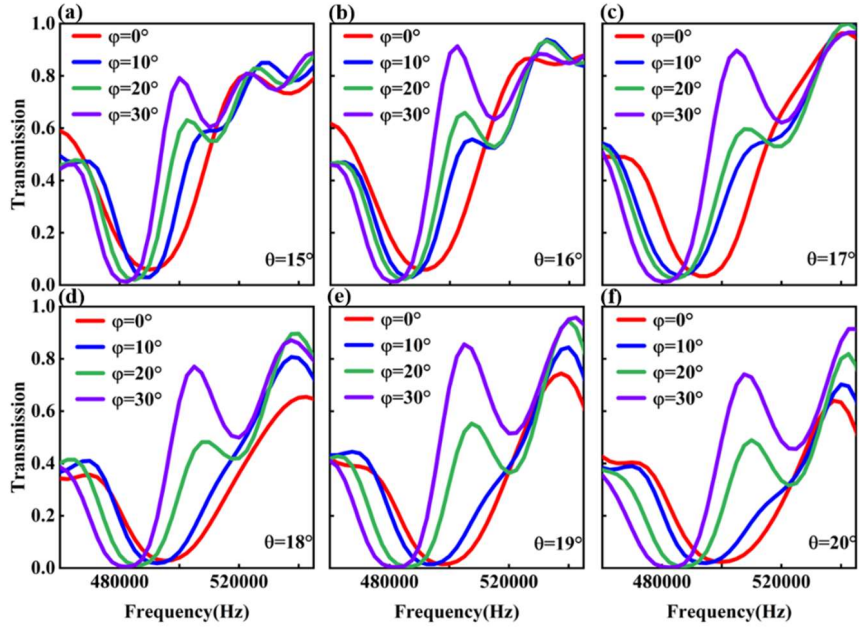

FIG. 5. The experimentally measured transmission spectra versus the frequency for different incidence angles.

Further, to demonstrate that the phononic slab exhibits a line of BICs, we repeat the experiment for different values of $\theta$ (from $15°$ to $20°$). As is shown in Fig. 5, in all cases, the Fano resonances associated with the quasi-BICs are clearly seen for large values of $\varphi$, which confirms that the BICs indeed form a continuous line. In fact, we also repeat the experiment for small values of $\theta$ ($<15°$), but the Fano resonances associated with the quasi-BICs are so sharp that they exceed the frequency resolution limit of our experimental apparatus and thus do not appear clearly in the transmission spectra.

In conclusion, we demonstrate that a simple phononic crystal slab can support lines of SP BICs at and off the Γ point. Group theory methods and mode expansion methods are exploited to provide a clear understanding of the formation of the BICs. Moreover, it is revealed that these BICs correspond to lines of phase singularities of the radiation field. An integral of phase on a closed loop around the dislocation line defines a conserved topological charge. Due to its topological nature, the proposed BICs system is versatile, robust, and easy to realize. Our findings will have implications for implementing multi-frequency and multi-wave vector applications using BICs.

**Acknowledgements**

The authors would like to thank Professor Jiuyang Lu for the fruitful discussions. This work was supported by the National Natural Science Foundation of China (Grants Nos. 12174353, 11974262, and 12122408), and the National Key R&D Program of China (Grant No. 2023YFB4603800).